\documentclass[printer]{aa}
\usepackage{graphics,epsfig,lscape}
\usepackage{natbib,aas_macros}
\bibpunct{(}{)}{;}{a}{}{,}

\def\cm-2{cm$^{-2}$}

\def\chandra{{\it Chandra~}}
\def\sax{{{\it Beppo}SAX~}}
\def\xmm{{XMM-{\it Newton}~}}
\def\n4258{\object{NGC~4258}}
\newcommand{\ergcm}[1]{$\times10^{#1}$ \hbox{erg cm$^{-2}$ s$^{-1}$}}
\newcommand{\oergcm}[1]{$10^{#1}$ erg cm$^{-2}$ s$^{-1}$}
\newcommand{\ergs}[1]{$\times10^{#1}$ \hbox{erg s$^{-1}$}}
\newcommand{\oergs}[1]{$10^{#1}$ erg s$^{-1}$}
\newcommand{\hcm}[1]{$\times10^{#1}$ cm$^{-2}$}
\newcommand{\ohcm}[1]{$10^{#1}$ cm$^{-2}$}
\newcommand{\expo}[1]{$\times10^{#1}$}

\newcommand{\nh}{\hbox{$N_{\rm H}$}}

\begin{document}

   \title{An X-ray view of the active nucleus in \n4258\thanks{Based 
    partly on observations with XMM-{\it Newton}, an ESA Science Mission 
    with instruments and contributions directly funded by ESA Member
    States and the USA (NASA).}}

   \author{W. Pietsch\inst{1} \and  
           A.M. Read\inst{1,2} 
          }
\institute{Max-Planck-Institut f\"ur extraterrestrische Physik, 85741 Garching, Germany 
    \and School of Physics \& Astronomy, University Birmingham, Birmingham B15 2TT, UK}
     
     \offprints{W.~Pietsch}
     \mail{wnp@mpe.mpg.de}

   \date{Received; accepted }

        \abstract{\xmm observed the Seyfert 1.9 galaxy \n4258 in
	December 2000. At energies above 2~keV a hard nuclear point
	source is resolved that can be fitted by a highly absorbed
	power-law spectrum (\nh = (8.0$\pm$0.4)\hcm{22}, photon index
	1.64$\pm$0.08) with an unabsorbed luminosity of 7.5\ergs{40}
	in the (2--10) keV band. No narrow iron K$\alpha$ emission
	line is detected (90\% upper limit of equivalent width EW
	$\sim$40 eV). The nuclear emission flux was observed to remain
	constant over the observation. A short archival \chandra
	observation taken in March 2000 further constrains the hard
	emission to a point source coincident with the radio
	nucleus. A point source $\sim$3\arcsec\ southwest of the
	nucleus does not contribute significantly. Spectral results of
	the \chandra nuclear source are comparable (within the limited
	statistics) to the \xmm parameters. The comparison of our iron line
	upper limit with reported detections indicates variability of the line 
	EW. These results can be explained by the relatively low nuclear
	absorption of \n4258 (which is in the range expected for its 
	intermediate Seyfert type) and some variability of the absorbing
	material. Reflection components as proposed to explain the large iron 
	line EW of highly absorbed Seyfert 2 galaxies  and/or variations 
	in the accretion disk are however imposed by the time variability of the
	iron line flux. 
} 
\maketitle

\section{Introduction}
\n4258\,(\object{M106}) is a B$_{\rm T}$ = 8.5 mag nearby
\citep[distance of 7.2 Mpc,
i.e. 1$"\cor35$~pc,][]{1999Natur.400..539H}, highly inclined
\citep[72\degr, ][]{1988ngc..book.....T} SABbc spiral
spectroscopically classified as a 1.9 Seyfert galaxy
\citep{1997ApJS..112..315H}.  The strong polarization of the
relatively broad optical emission lines \citep{1995ApJ...455L..13W}
further supports the existence of an obscured active nucleus
in \n4258. Water maser line emission of rotating gas near the center
of \n4258 is modeled by a 3.6\expo{7} M$_{\sun}$ mass within 0.13 pc
of the nucleus, indicating a massive nuclear black hole inside a
highly inclined (83\degr) thin gaseous disk
\citep{1995Natur.373..127M,1995ApJ...440..619G,1995A&A...304...21G}.

\citet{1994ApJ...421..122T} detected a very bright radio continuum
nuclear source, showing, in observations from March 1985 through to
May 1990, long term time variability by 47\% both at 6 and 20 cm 
\citep{2001ApJ...551..702H}.

In X-rays ROSAT PSPC and HRI observations of \n4258
\citep{1994A&A...284..386P,1995ApJ...440..181C,1999A&A...352...64V}
resolved diffuse emission mostly connected to the ``anomalous'' arms
and emission from \n4258 point sources. However, due to the soft
energy band (0.1--2.4 keV) only upper limits could be derived for
emission from the nucleus.  ASCA observations of \n4258 were of
reduced spatial resolution but extended the energy coverage to 10 keV
with improved spectral resolution. \citet{1994PASJ...46L..77M}
reported, as well as soft emission components, a hard, highly absorbed
point-like component that could be modeled by an absorbed power-law
(\nh\ $\sim1.5$\hcm{23}, photon index $\sim$1.78, unabsorbed luminosity
4\ergs{40} in the (2--10) keV band) and an iron K emission line with
an equivalent width of (250$\pm$100) eV. Further ASCA observations
improved on the spectral parameters of the hard component, confirming
a narrow iron K$\alpha$ emission line at ($6.45^{+0.10}_{-0.07}$) keV
with an equivalent width of ($107^{+42}_{-37}$) eV, and showing time
variability of the (5--10) keV flux and probably also the absorbing
column of the hard component \citep{2000ApJ...540..143R}.  \sax
observations extended the spectral coverage of the power-law component
to 70 keV \citep{2001ApJ...556..150F}. \citet{2001ApJ...560..1W} have
recently reported the results of a \chandra observation optimized to
investigating the X-ray emission from the ``anomalous arms". The
region around the nucleus showed two compact sources, a bright heavily
absorbed source coincident with the radio nucleus and the nuclear
H$_2$O maser source, and a second fainter source offset to the SW by
2\farcs5. While the nuclear source is heavily piled-up, debilitating
the determination of reliable spectral parameters, the spectrum of the
weaker source is well described by an absorbed power-law
(\nh\ $\sim2$\hcm{21}, photon index $\sim$1.5, unabsorbed luminosity
5\ergs{38} in the (0.5--4.5) keV band), and is most likely an X-ray
binary within \n4258 itself.

An \xmm observation of \n4258 was carried out to investigate the
different emission components.  In this paper we will concentrate on
the hard X-ray emission from the active nucleus and defer an analysis
of the \n4258 diffuse emission and extra-nuclear point sources to a
later paper.  We also analyzed a short \chandra ACIS-S observation
from the \chandra archive adding to the nuclear point source
interpretation of the hard emission.

\section{\xmm observations and results}
\begin{figure}
   \resizebox{\hsize}{!}{\includegraphics[bb=70 80 500 510,clip]{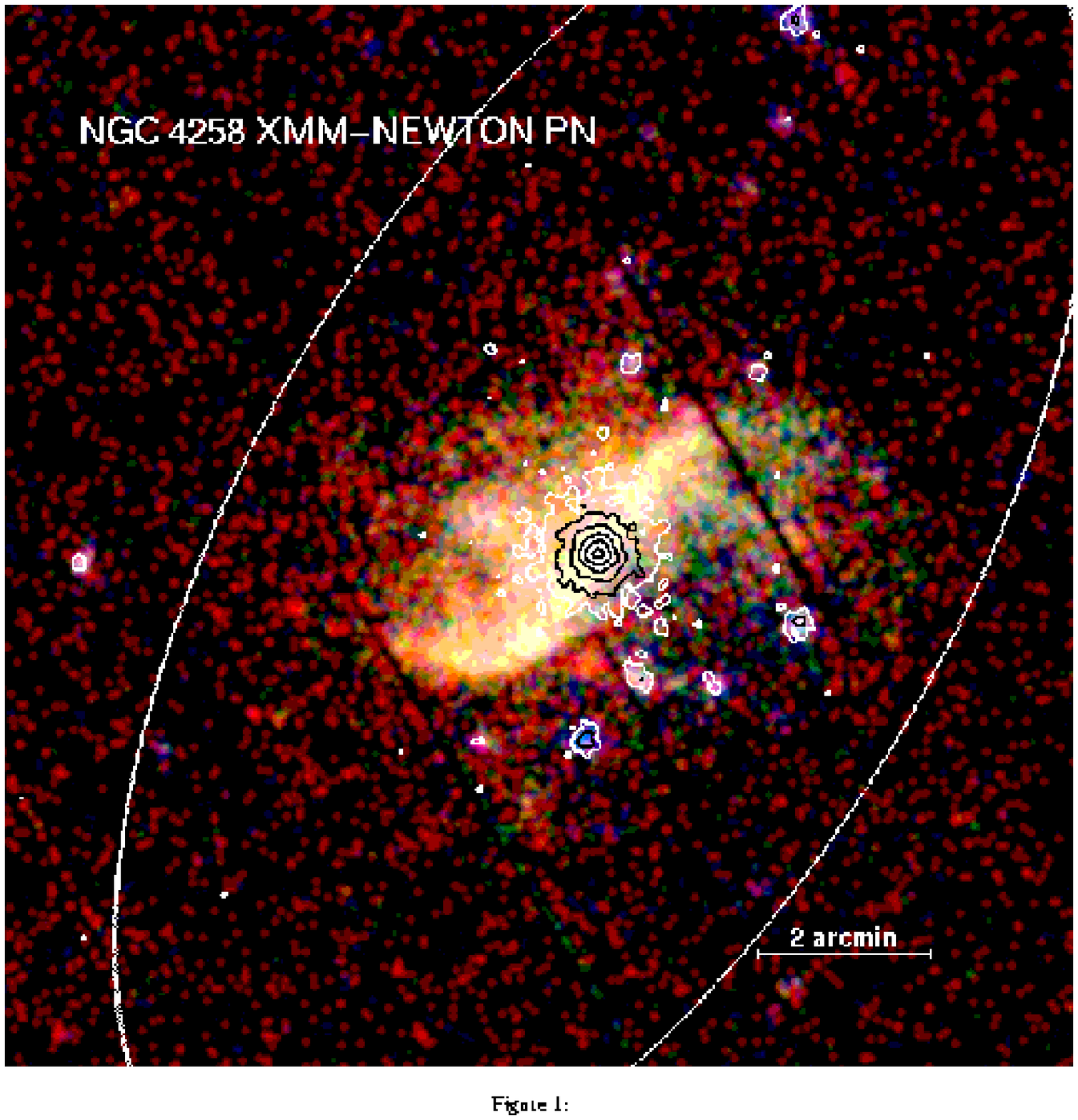}}
    \caption[]{
Logarithmically-scaled, three-color \xmm EPIC PN image of the \n4258 disk
and nuclear regions. Red, green and
blue show respectively the ROSAT-equivalent (0.2--0.5)~keV, (0.5--0.9)~keV 
and (0.9--2.0)~keV bands, while the hard
(2--10)~keV emission is shown superimposed as black/white 
contours at levels increasing by factors of 3 from 0.12\,ct~arcsec$^{-2}$. 
The data in each energy band have been smoothed with a PSF-equivalent Gaussian
of FWHM 5\arcsec. The inclination-corrected optical D$_{25}$ ellipse
of \n4258 is marked
    \label{color}}
\end{figure}

\begin{figure}
  \resizebox{\hsize}{!}{\includegraphics[angle=-90]{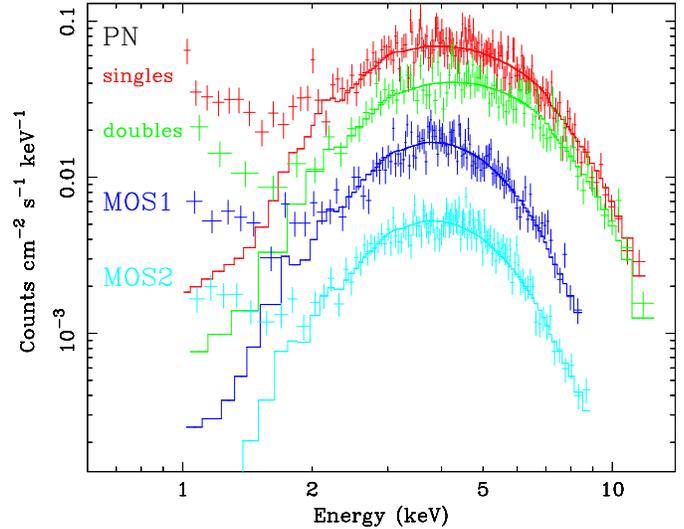}}
    \caption[]{\xmm EPIC background-subtracted spectra of 
    the nuclear area of \n4258 (extraction radius 15") for 
    energies above 1 keV integrated during low 
    background times with an absorbed power-law model (see
    text) indicated. PN singles, PN doubles, 
    MOS1, and MOS2 data were fitted simultaneously for energies above 2 keV. 
    MOS1 and MOS2 intensities were shifted down for clarity by 0.5 
    and 1 decade, respectively
    }
     \label{epic_spectra}
\end{figure}

\begin{figure}
  \resizebox{\hsize}{!}{\includegraphics[angle=-90]{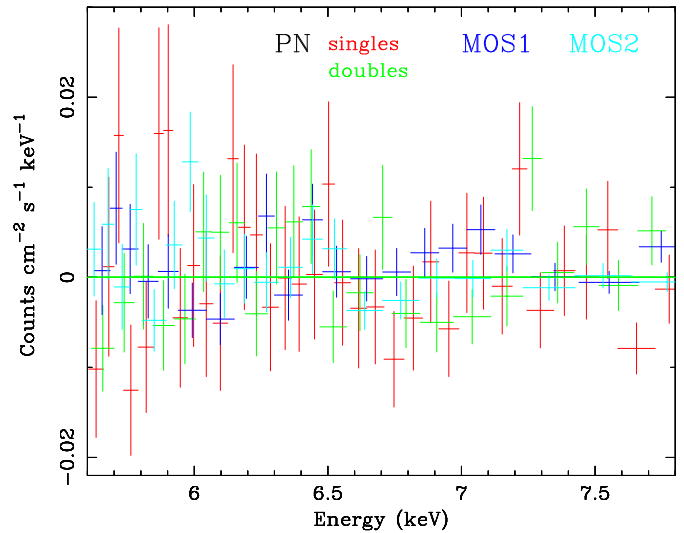}}
    \caption[]{Residuals of data versus power-law model of the spectral fit
    to the nuclear area of \n4258 (see Fig.~\ref{epic_spectra}) in the
    energy band around the iron K$\alpha$ line energy (5.5--7.8 keV) for EPIC PN 
    singles and doubles, MOS1 and MOS2, respectively
     }
     \label{epic_res}
\end{figure}

\n4258 was observed with \xmm \citep{2001A&A...365L...1J} on December
8/9, 2000 during orbit 183. The observations with the EPIC PN
\citep{2001A&A...365L..18S} and MOS \citep{2001A&A...365L..27T}
detectors were of $\sim$16.5 ks and $\sim$20.6 ks respectively, with
the PN in Extended Full Frame mode (thin filter) and MOS1 and MOS2
both in Full Frame mode (medium filter).  Calibrated EPIC event files
were created with the \xmm SAS (Science Analysis Software) version
20010512\_0642.  X-ray events corresponding to patterns 0--4 (singles
and doubles) were then selected for the PN camera for spectra and
images; for the MOS cameras patterns 0--12 were used. Known hot or bad
pixels were removed during screening. High background times at the end
of the observation were rejected from the spectral fitting analysis,
reducing usable integration times for spectra to 13.8, 19.9, and 20.0
ks for PN, MOS1 and MOS2, respectively.

An EPIC PN color image of the \n4258 nucleus and inner disk area was
created with contours of the 2--10 keV emission superimposed
(Fig.~\ref{color}).  Red, green and blue show respectively the
ROSAT-equivalent (0.2--0.5)~keV, (0.5--0.9)~keV and (0.9--2.0)~keV
bands. The image is dominated by complex unresolved emission in the
soft bands with an extent of at least 6\arcmin\ and a bright nuclear
point source in the hard band.  In addition there are several fainter
point-like sources distributed over the \n4258 disk.

To characterize the nuclear emission, we extracted PN spectra from
different extraction radii separately for single events and double
events.  For MOS1 and MOS2 spectra, we added together single, double,
triple and quadruple events.  As expected from Fig.~\ref{color}, the
contribution of the extended emission to the soft band strongly
decreases for smaller extraction radii. Due to the different spectral
shape, the soft contributions clearly separate from the harder nuclear
emission and do not contribute above $\sim$2 keV. We therefore
restricted our spectral modeling to energies above this threshold.
Figure~\ref{epic_spectra} shows spectra integrated for the different
EPIC detectors and pattern selections using a source extraction radius
of 15\arcsec\ which covers $\sim$80\% of the integrated point spread
function PSF \citep{2000SPIE.4012..731A}, and background spectra from
a 50\arcsec radius area free of extended emission $\sim$4\farcm2 to
the NE of the nucleus. For clarity, the MOS1 and MOS2 intensities have
been shifted down by 0.5 and 1 decade respectively. Spectra were
binned to a minimum of 30 counts per bin in order to have sufficient
statistics in the individual bins for $\chi^2$ minimum fitting.  For
spectral fits we used the most recent response matrices for PN singles
and doubles (epn\_ff20\_sY9\_thin.rmf and epn\_ff20\_dY9\_thin.rmf,
Mar 2001) and MOS1 and MOS2 (m1\_medv9q19t5r4\_all\_15.rsp and
m2\_medv9q19t5r4\_all\_15.rsp, Feb 2001). If not quoted otherwise
errors are at the 90\% confidence level (e.g.  $\Delta\chi^2 = 2.7$
for one interesting parameter).

A highly absorbed power-law model (\nh\ = (8.0$\pm$0.4)\hcm{22},
photon index $\Gamma$ = 1.64$\pm$0.08) yielded an acceptable fit
(reduced $\chi^{2} = 0.98$ for 336 degrees of freedom, dof; see
Fig.~\ref{epic_spectra}). The normalization factors of 1.10$\pm$0.04
and 1.14$\pm$0.04 between MOS1 and PN and MOS2 and PN respectively,
are caused by differences in the PSF of the instruments as well as
absolute calibration uncertainties. The resulting absorbed and
unabsorbed fluxes in the hard (2--10 keV) band are 7.6 and
12.1\ergcm{-12} respectively, corresponding to luminosities of 4.7 and
7.5\ergs{40}. As one might expect due to the limited energy range, a
hard thermal bremsstrahlung model was equally acceptable (\nh\ =
7.4\hcm{22}, $kT = 21.7$ keV, reduced $\chi^{2} = 0.95$ for 336 dof).

The residuals of the data versus the power-law model in the energy
band around the iron K$\alpha$ line energy (5.5--7.8 keV, see 
Fig.~\ref{epic_res}) do not
support any emission line in this energy range, specifically not a
narrow iron line around 6.45 keV as reported by ASCA. From the \xmm
data we determined the 90\% upper limit for the photon flux of a
narrow emission line around 6.45 keV to be 4.2$\times$10$^{-6}$
photons~cm$^{-2}$~s$^{-1}$ (corresponding to an equivalent width of
$\sim$40~eV). This upper limit is much lower than the values reported
from the ASCA detections.

We searched for time variability of the nuclear source in the 2--10
keV band.  To do so we extracted EPIC light curves of the nuclear
source with a time resolution of 200 s. There is no long term trend
within the observation.  We also find no flux changes greater than
$\sim$20\% from the observation average on shorter time scales down to
our sampling of 200 s.

\section{\chandra observations and results}

\begin{figure}
  \resizebox{7.8cm}{!}{\includegraphics[bb=84 109 522 501,clip]{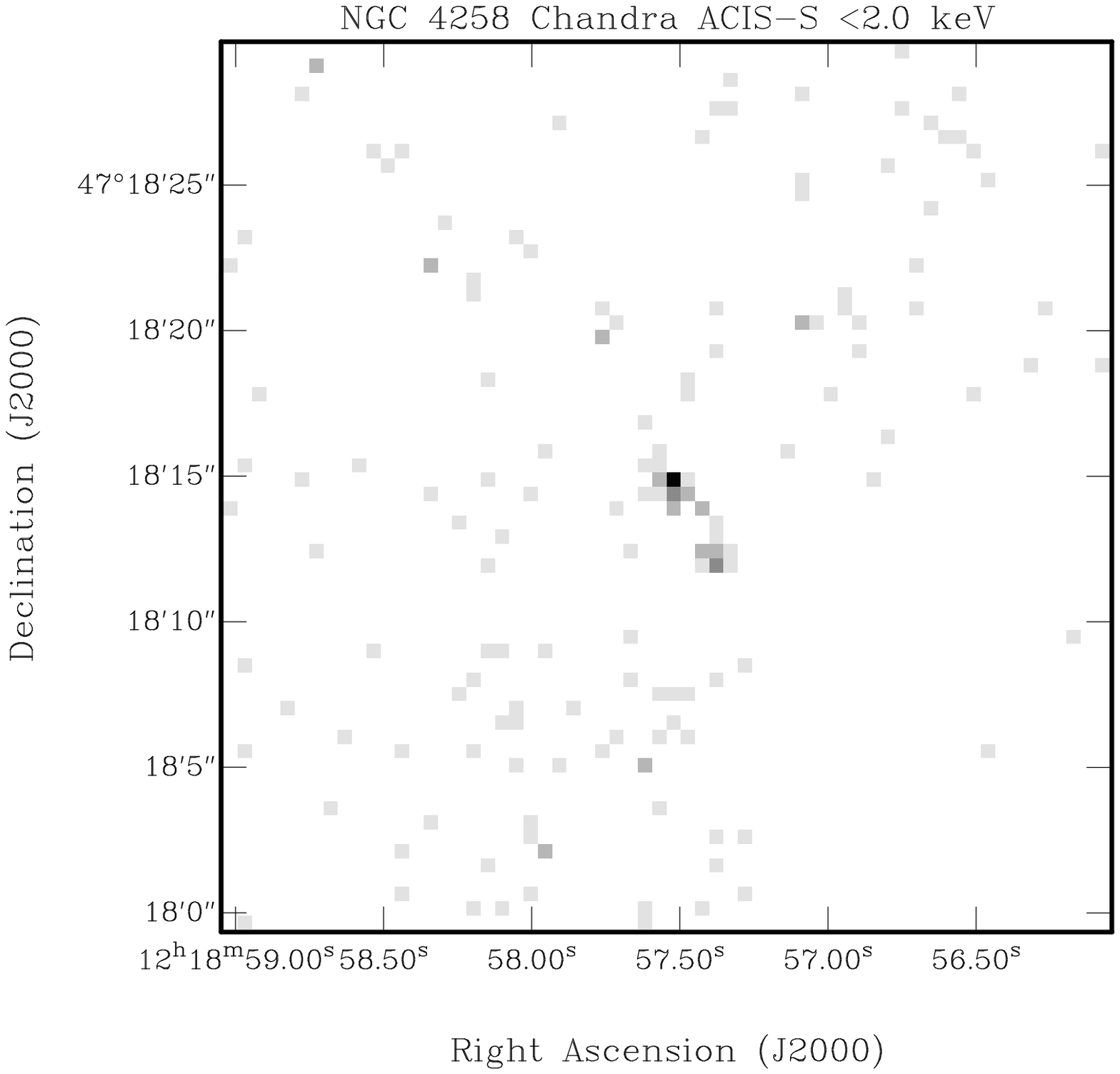}}
  \resizebox{7.8cm}{!}{\includegraphics[bb=84 94 522 510,clip]{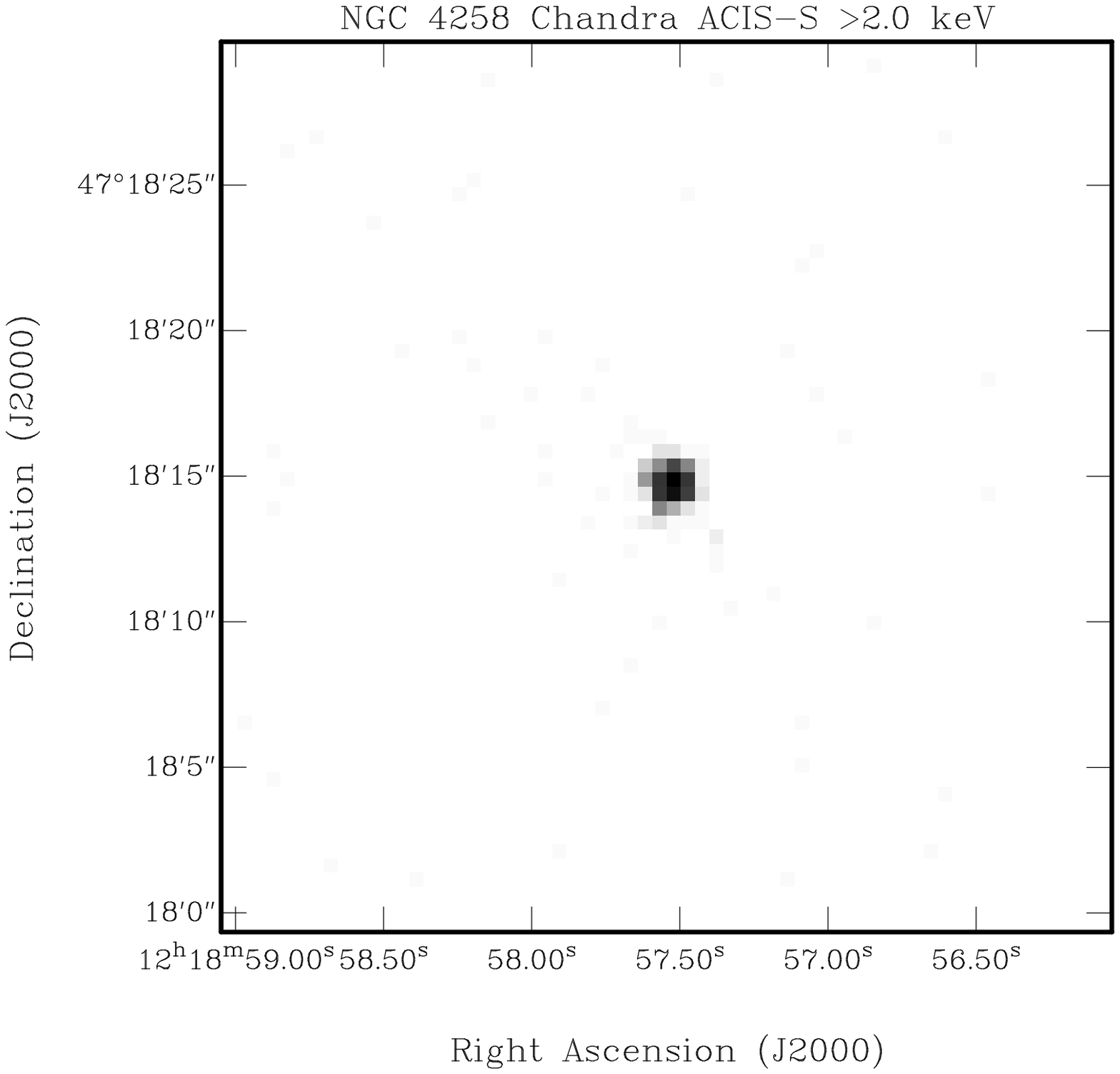}}
    \caption[]{
  Linearly-scaled, grey-scale \chandra ACIS-S images of the \n4258 nuclear region.
  The images (RA, Dec J2000.0) were binned with a pixel size of 0\farcs5 in the 
  soft ($<$2 keV, top) and hard ($>$2 keV, bottom) band and have maxima of 6 
  and 109 counts per pixel respectively 
    }
    \label{chandra_image}
\end{figure}
\begin{figure}
  \resizebox{\hsize}{!}{\includegraphics[angle=-90]{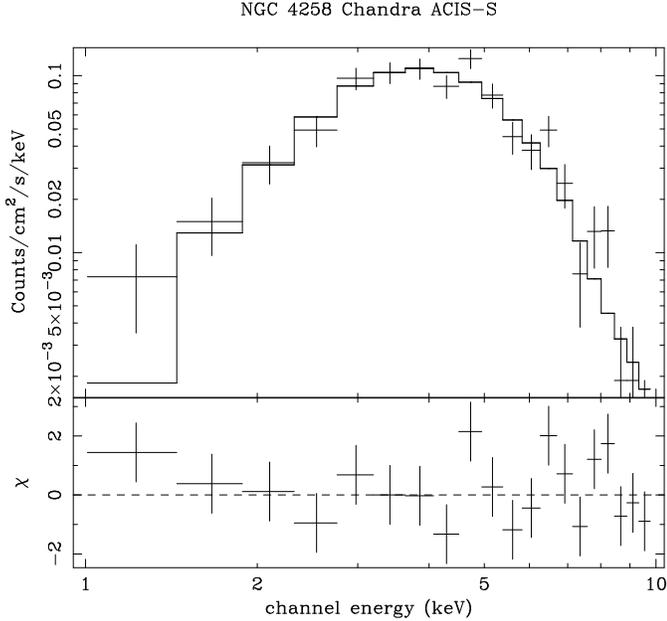}}
    \caption[]{\chandra background-subtracted spectrum of the nuclear area of 
    \n4258 (extraction radius 2\farcs0625) with
    an absorbed power-law model (see text) 
    indicated 
    }
     \label{chandra_spectrum}
\end{figure}

Data (cleaned event files, images and aspect offset files) from a
short exposure \chandra ACIS-S observation of NGC~4258, performed in
March 2000, were obtained from the \chandra Data Archive
(http://asc.harvard.edu/cgi-gen/cda).  No flaring in the data was
observed, and so the data over the full integration time (1205\,s)
were used to create images and spectra.

Figure~\ref{chandra_image} shows raw ACIS-S images in the (top) soft
band (below 2\,keV) and the (bottom) hard band (above 2\,keV). The
images are 30\arcsec\ to a side, equivalent to the 30\arcsec\ diameter
extraction circles used in the creation of the XMM EPIC NGC~4258
nuclear spectra.

A dominant source is detected at $\alpha$=12$^{\rm h}$18$^{\rm
m}$57\fs53, $\delta$=+47\degr18\arcmin14\farcs7 (J2000), lying
0\farcs7 from the NGC~4258 central radio source of
\citet{1994ApJ...421..122T}, and is very prominent in the hard band
image. Though it is also visible in the soft band image (though less
so), what is most striking is the appearance of a soft companion
$\approx$3\arcsec\ to the southwest. This companion is just about 
a factor of 2 fainter than the nuclear source in the soft band, whereas 
in the hard band it is fainter by a factor of more than 100.

As this short Chandra observation had not been optimised to study the
``anomalous arms'' (as had the longer observation presented in 
\citet{2001ApJ...560..1W}), no significant pile-up is seen from the 
nuclear source,
and the \chandra spectrum from the \n4258 nucleus can be studied for
the first time.
Programs within the \chandra data analysis suite CIAO were used to
extract a spectrum from a 2\farcs0625 radius circle about the position
of the bright source. Similarly, a background spectrum was extracted
from a source-free 9\farcs375 radius circle some 13\arcsec\ away to
the northeast. Once response matrix files and ancillary response files
were created (again using CIAO), the spectra were binned to a minimum
of 30 counts per bin. Fitting of an absorbed power-law model to the
data between 1 and 10\,keV (see Fig.~\ref{chandra_spectrum}) gives
rise to results entirely consistent with those obtained with XMM EPIC:
\nh\ = (7.2$\pm$1.8)\hcm{22}, $\Gamma$ = 1.4$\pm$0.5, with a
reduced $\chi^{2}$ of 1.36. The large errors are due to the poor
statistics obtained over such a short integration time. Note the
$\sim$1$\sigma$ suggestion of an iron-like feature at around 6.5\,keV.

Assumption of this model gives rise to absorbed and unabsorbed hard
band, (2--10) keV fluxes of 1.44 and 2.12\ergcm{-11} respectively,
corresponding to luminosities of 0.89 and 1.32\ergs{41}.

\section{Discussion}

\begin{table*}
\caption[]{Historical high energy spectral data for \n4258.}
\begin{tabular}{lrrrrrccr}
\hline\noalign{\smallskip}
\multicolumn{1}{l}{Observatory}  & \multicolumn{1}{c}{Date} &
\multicolumn{1}{c}{\nh} & \multicolumn{1}{c}{$\Gamma$} &
 \multicolumn{1}{c}{$E_{\rm line}$} & \multicolumn{1}{c}{$W_{\rm line}$} &
\multicolumn{1}{c}{$f_{\rm X}^{ *}$}    & \multicolumn{1}{c}{$L_{\rm X}^{ **}$}    & 
\multicolumn{1}{c}{Ref.}\\ 
\noalign{\smallskip}
& & \multicolumn{1}{c}{[\ohcm{22}]} &   & 
\multicolumn{1}{c}{[keV]} & \multicolumn{1}{c}{[eV]} &  &   & \\
\noalign{\smallskip}\hline\noalign{\smallskip}
ASCA     & May 1993 & $15\pm2$             & $1.78\pm0.29$          & $6.5\pm0.2$ & $250\pm100$ &     & 4.2  &  1   \\
ASCA     & May 1993 & $13.6^{+2.1}_{-2.2}$ & $1.78^{+0.22}_{-0.26}$ &             &             & 5.1 &      &  2   \\
ASCA     & May 1996 & $9.2\pm0.9$          & $1.71^{+0.18}_{-0.17}$ &             &             & 8.3 &      &  2   \\
ASCA     & Jun 1996 & $8.8^{+0.7}_{-0.6}$  & $1.83\pm0.13$          &             &             & 8.8 &      &  2   \\
ASCA     & Dec 1996 & $9.7\pm0.8$          & $1.87\pm0.15$          &             &             & 9.5 &      &  2   \\
\sax     & Dec 1998 & $9.4\pm1.2$          & $2.11\pm0.14$          & $6.57\pm0.20$ & $85\pm65$ &     & 10   &  3   \\
ASCA     & May 1999 & $9.5^{+2.1}_{-0.9}$  & $1.86^{+0.40}_{-0.13}$ &             &             & 4.0 &      &  2   \\
ASCA     & May 1999 & $8.9^{+0.4}_{-0.7}$  & $1.83^{+0.06}_{-0.09}$ & $6.45^{+0.10}_{-0.07}$ &  $107^{+42}_{-37}$ &  & 5.7 &  2   \\
\chandra & Mar 2000 & $7.2\pm1.8$          & $1.4\pm0.5$            &             &             &     & 13   & this work \\
\xmm     & Dec 2000 & $8.0\pm0.4$          & $1.64\pm0.08$          &  6.45       & $<40$       &     &  7.5 & this work \\
\noalign{\smallskip}
\hline
\noalign{\smallskip}
\end{tabular}
\label{spectra}

Notes and references:\\
$^{ *~}$: 5--10 keV uncorrected flux [\oergcm{-12}]\\
$^{ **}$: 2--10 keV absorption corrected luminosity for \n4258 distance 
of 7.2 Mpc [\oergs{40}] \\
(1) \citet{1994PASJ...46L..77M};
(2) \citet{2000ApJ...540..143R};
(3) \citet{2001ApJ...556..150F}
\end{table*}

ASCA observations of \n4258 established a heavily absorbed hard
component in the overall X-ray spectrum which was attributed to the
nucleus and confirmed by \sax \citep{1994PASJ...46L..77M,
2001ApJ...556..150F}. The ASCA and \sax observations of the hard
emission of \n4258 however, suffered from instrument PSFs of more than
an arcminute, and only higher resolution instruments such as \xmm and
\chandra were thought able to confirm a nuclear origin to the hard
emission.  Even the \xmm observations do not resolve the nuclear
emission from that of another reasonably bright nearby point source,
most likely an X-ray binary, that is resolved by \chandra (see
Sect. 3).  \citet{2001ApJ...560..1W}, analysing a deeper \chandra
observation, report this source to be at a projected distance from the
nucleus of 2\farcs5 (87 pc) and characterize its spectrum as an
absorbed power-law ($N_{\rm H} = (2.0^{+1.2}_{-1.1})$\hcm{21}, $\Gamma
= 1.49^{+0.50}_{-0.37}$, absorption corrected luminosity
5.1\ergs{38}), similar in slope but much less absorbed than the
nuclear spectrum.  Fortunately, the source is rather faint (less than
1\% in flux) compared to the nuclear emission and cannot significantly
influence the \xmm results. Also we detect no other hard bright
point-like sources within an arcmin of the nucleus that could
significantly contribute to the ASCA and \sax emission, which
therefore really has to originate from the nucleus.

In Table \ref{spectra} we give an historical record of the \n4258
nuclear spectrum. The \xmm power-law spectral parameters are within
the range of the other measurements. The photon index is on the hard
side but compatible within the errors of the ASCA results. Only the
\sax index pointed at a somewhat softer spectrum. The absorbing column
($\sim$8\hcm{22}) seems to have not varied significantly since the
first ASCA measurements that pointed at an absorption a factor 1.5--2
higher. The luminosity however, seems to have varied on a timescale of
several years by a factor of about two starting from low values in
1993 with a maximum in 1996 to 1998. The \xmm measurements indicate
that in December 2000 the nuclear luminosity was again at a lower
level. \citet{2001ApJ...556..150F} report variability of a factor of
about 2 on half-day timescales as well as smaller scale variations
(10\%--20\%) on timescales as short as 1 hr. During the \xmm
observations the source did not show any variability at similar
amplitudes. With its absorption-corrected (2--10) keV luminosity of
typically 7\ergs{40}, the AGN in \n4258 is presently not very active
compared to other Seyfert galaxies. \citet{1999ApJ...522..157R} 
report luminosities from  Seyfert galaxies ranging from  3\ergs{40} to 
2\ergs{44}.

The high X-ray absorption of the nuclear component in \n4258 is
expected, given the optically-derived Seyfert 1.9 classification of
the galaxy, using the so-called ``unified model''
\citep{1993ARA&A..31..473A}. This model assumes that the engine is at
work in all active galactic nuclei and differences between type 1 and
type 2 AGN are ascribed solely to orientation effects, i.e. that our
line of sight is (type 2) or is not (type 1) obscured by optically
thick material.  \citet{1999ApJ...522..157R} showed that most ``strict''
Seyfert 2 nuclei are highly obscured (\nh $>$\ohcm{23}) or even
Compton thick(\nh $>$\ohcm{24}). Intermediate type 1.8--1.9 Seyfert
galaxies on the other hand are characterized by an average \nh\ much
lower than ``strict'' Seyfert 2 galaxies. The absorption of the \n4258
nuclear spectrum of $\sim$8\hcm{22} fits into this scenario.

From the first reports of hard X-ray emission from \n4258, a narrow
iron K$\alpha$ emission line was always needed to model the
spectra. According to Table~\ref{spectra}, the strongest line compared
to the continuum was needed for the 1993 ASCA data. With \xmm we do
not detect a line, and our EW upper limit is more than a factor of 6
below the \citet{1994PASJ...46L..77M} initial ASCA detection and still
a factor of more than 2 below the later ASCA and \sax values
\citep{2000ApJ...540..143R, 2001ApJ...556..150F}. Note that
\citet{1994PASJ...46L..77M} report that no background was subtracted
from the May 1993 ASCA raw spectra and therefore, one might have
doubts regarding the line strength in this analysis if the background
contains contributions from an intrinsic iron feature. It must be
stated however, that the existence of the line was clearly
demonstrated in the later ASCA observations. The \sax equivalent width
from December 1998 is, due to the large errors, compatible with both
the ASCA results from May 1999 and with our \xmm upper limit from
December 2000.

There are at least two possible explanations for this result: (1) The
difference could be caused by the larger sky area used to accumulate
the ASCA and \sax spectra. In this way an extended component from the
\n4258 disk could be picked up that would not be covered by the much
smaller spectral extraction region used here with \xmm. We tried to
rule out this explanation by extracting \xmm spectra from a comparably
large area. In these spectra we did not detect any iron line
emission. The upper limit to the equivalent width however is similar
to the measurements by earlier satellites. (2) The most
straightforward explanation however, is that the line intensity is
varying with time.

Several groups have calculated the contribution of an obscuring torus
to the X-ray spectra of Seyfert galaxies using Monte Carlo simulations
\citep[e.g.][]{1991PASJ...43..195A, 1993MNRAS.263..314L,
1994MNRAS.267..743G}. They find strong dependence of the iron line EW
at 6.4 keV on the absorbing column and geometrical factors. For the
relatively low nuclear absorption in \n4258, EWs of $\sim$100 eV or
less are expected. This is compatible with our findings. These simple
models therefore would be sufficient to explain the \xmm data 
and we would not need
to invoke additional reflection components. Such a component is for
instance needed to explain the iron line EW of $>$2 keV measured by
\chandra in the Seyfert 2 nucleus of M\,51
\citep{2001ApJ...560..139T}.

According to Table \ref{spectra}, the iron line varied in absolute flux
by a factor of at least 2 in 18 months (from May 1999 to December 2000) setting
an upper limit on the size-scale of the fluorescing region of 0.5 pc. This
seems to argue against a torus origin for the iron line which would imply
a significantly larger size scale. Instead, it favours an origin of the line 
in the accretion disk much closer to the nucleus. If the line originated
in the outer regions of the accretion disk as proposed by 
\citet{2000ApJ...540..143R} using different arguments, its variability could 
reflect intensity changes of the nuclear power law source (that may not have 
been covered directly by the observations listed in Table \ref{spectra}). 
Alternative explanations for the line variability might
be changes in the disk (e.g. the ``inner'' edge of the optically thin disk, or
the ionization structure of the disk surface).

The - up till now - sparse data on the \n4258 6.4 keV line seem to
indicate a correlation between the amount of absorption and the EW of
the line. However, an anticorrelation of the line EW with the nuclear
luminosity can also not be ruled out. More high quality spectra are
clearly needed to clarify the behaviour, and such results may then
force revisions and improvements to the simple unification picture.

\section{Summary}
The \xmm observations of the Seyfert 1.9 galaxy \n4258 show a hard nuclear 
point source with a spectrum that can be modeled by a highly absorbed 
power-law. The analysis of a short archival \chandra observation constrains the
hard emission to a point source coincident with the radio nucleus.
During the \xmm observation the nuclear emission flux remained constant.
Our upper limit to narrow iron K$\alpha$ line emission is significantly lower
than earlier detections by ASCA and \sax and indicates time variability of this
component. Simple absortion models would be sufficient to explain the reported line 
fluxes and upper limits. The suggested long term line variability, however,
seems to point at a more complex scenario including reflection from the
accretion disk and/or variability of the disk geometry and ionization.

The detailed analysis of \xmm data from the extranuclear point sources 
as well as from the diffuse emission from the ``anomalous'' arms and the 
interstellar medium in the disk and halo of \n4258  will be published in a 
following paper. 
  
\begin{acknowledgements}
    We thank the referee (Christopher Reynolds) for helpful comments. 
    The \xmm project is supported by the Bundesministerium f\"{u}r
    Bildung und Forschung / Deutsches Zentrum f\"{u}r Luft- und Raumfahrt 
    (BMBF/DLR), the Max-Planck Society and the Heidenhain-Stiftung.
\end{acknowledgements}
\bibliographystyle{apj}
\bibliography{./paper,/home/wnp/data1/papers/my2000,/home/wnp/data1/papers/my2001}

\end{document}